\documentclass[twocolumn,showpacs,preprintnumbers,amsmath,superscriptaddress,amssymb,prb]{revtex4-2}
\usepackage{dcolumn}% Align table columns on decimal point
\usepackage{bm}% bold math
\usepackage[dvipdfmx,hidelinks]{hyperref}
\newcommand{\doijournal}[2]{\href{https://doi.org/#1}{#2}}
\usepackage[dvipdfmx]{graphicx}

\begin{document}

%\preprint{APS/123-QED}

\title{Nonlocal current-response theory of structured-light dichroism}

\author{Akihito Kato}
 \affiliation{Department of Physics and Electronics, Osaka Metropolitan University, 1-1 Gakuen-cho, Sakai, Osaka 599-8531, Japan}
 \affiliation{Institute for Molecular Science, National Institutes of Natural Sciences, Okazaki 444-8585, Japan}
\author{Nobuhiko Yokoshi}
 \affiliation{Department of Physics and Electronics, Osaka Metropolitan University, 1-1 Gakuen-cho, Sakai, Osaka 599-8531, Japan}

\date{\today}% It is always \today, today,

\begin{abstract}
We develop a nonlocal minimal-coupling theory of structured-light dichroism.
Optical absorption is written as a bilinear functional of the incident vector potential and the nonlocal current-response kernel, retaining the spatial structure of optical vortex beams and other inhomogeneous fields.
Circular dichroism, helical dichroism, and helical circular dichroism are treated as distinct reversal channels that project different reversal-odd components of the same microscopic response.
The resulting signal is resolved into symmetry, tensor, and angular-mode sectors.
Single azimuthal-mode fields probe diagonal OAM-channel components, whereas mixed modes access off-diagonal mode-space coherence through interference between different OAM channels.
The relative polarization of the mixed field further selects scalar, axial-vector, or rank-2 tensor sectors of the response.
This decomposition gives selection rules for high-symmetry systems and a diagnostic scheme for low-symmetry nanophotonic structures, finite systems, and extended materials, where several angular-channel responses may coexist.
It also connects the nonlocal current-response formulation to local-gradient descriptions based on spatial dispersion, optical chirality, and tensorial anisotropy.
\end{abstract}

\maketitle

%/_/_/_/_/_/_/_/_/_/_/_/_/_/_/_/_/_/_/_/_/_/_/_
\section{Introduction}
%/_/_/_/_/_/_/_/_/_/_/_/_/_/_/_/_/_/_/_/_/_/_/_

Light carrying angular momentum provides a powerful probe of internal degrees of freedom in matter.
In addition to circular polarization, structured optical fields can carry spatial angular structure through an azimuthal phase factor $e^{i\ell\phi}$ \cite{Allen1992PRA}.
Optical vortex beams therefore enable absorption processes beyond the plane-wave picture, including excitation of electronic, excitonic, and collective modes with spatially structured transition currents.
For such fields, dichroism can be defined not only by reversing the circular-polarization label, as in conventional circular dichroism spectroscopy \cite{RodgerNorden1997,Berova2000Book,Barron2004Book}, but also by reversing the azimuthal-mode index or by reversing both labels.
We call these operational reversal channels circular dichroism (CD), helical dichroism (HD), and helical circular dichroism (HCD), respectively.
Here CD denotes circular-polarization reversal, HD denotes azimuthal-mode-index reversal at fixed polarization, and HCD denotes simultaneous reversal of the circular-polarization and azimuthal-mode labels.
Structured-light dichroism has been studied in chiral media, molecular systems, nanostructures, metasurfaces, and nonlinear optical settings \cite{Loffler2011PRA,Brullot2016SciAdv,ForbesAndrews2021JPhysPhoton,ForbesJones2021PRA,Ni2021ACSNano,Begin2023NatPhoton,Jain2023JCP,ForbesGreen2024LPR,Jain2024ACSNano,Cheeseman2025AdvOptMater,Forbes2026PhotonicsResearch,Hashiyada2026Optica}.
In realistic beams, however, the optical field is often not a single ideal azimuthal-mode channel: Gaussian admixtures, nonparaxial corrections, and spin--orbit coupling can mix polarization and spatial angular degrees of freedom \cite{Marrucci2006PRL,Zhao2007PRL,Bliokh2010PRA,Bliokh2011OE,Forbes2026RPP}.

A microscopic description of these effects must retain the spatial structure of both the optical field and the material current response.
The minimal-coupling Hamiltonian provides a natural starting point because the vector potential couples directly to the electronic current density.
It gives a current-based formulation of absorption in which the optical mode profile is kept explicit \cite{List2015JCP,Foglia2022JCP,Bonafe2025PRB}.
This formulation is complementary to local multipolar descriptions, which are useful in the long-wavelength limit but can become cumbersome for strongly inhomogeneous fields or when many multipole orders contribute.
Spatial dispersion and nonlocal current-response functions therefore provide a natural language for structured and multipolar light--matter coupling \cite{Mukamel1995Book,Chernyak2015JCP,Lazzeretti2019JCP}.
They also allow pure azimuthal modes and mode superpositions to be treated on the same footing.

In this paper, we develop a nonlocal minimal-coupling theory of structured-light dichroism.
The absorption is written as a bilinear functional of the incident vector potential and the nonlocal current-response kernel, extending minimal-coupling analyses of molecular HD \cite{YeEtAl2019JCTC,JiangEtAl2023ChemSci}.
Within this framework, CD, HD, and HCD are reversal-odd projections of the same microscopic response.
The resulting signal is decomposed into symmetry, tensor, and mode-space sectors: scalar, axial-vector, and rank-2 material channels are separated, single azimuthal modes probe diagonal angular channels, and mixed modes probe off-diagonal coherence between different OAM-mode channels.
This decomposition gives selection rules in high-symmetry systems and a diagnostic scheme for low-symmetry nanophotonic structures, finite systems, and extended materials.
The theory thus provides a unified microscopic framework for interpreting structured-light dichroism beyond the plane-wave and purely local descriptions.

%/_/_/_/_/_/_/_/_/_/_/_/_/_/_/_/_/_/_/_/_/_/_/_
\section{Minimal nonlocal framework}
\label{sec.minimal}
%/_/_/_/_/_/_/_/_/_/_/_/_/_/_/_/_/_/_/_/_/_/_/_

We use the three operational reversal channels introduced above.
CD reverses the circular-polarization label $\sigma$ at fixed azimuthal-mode index $\ell$;
HD reverses $\ell$ at fixed $\sigma$;
and HCD reverses both labels $(\sigma,\ell)$.
Here $\sigma=\pm1$ labels circular polarization, and $\ell\in\mathbb Z$ labels the azimuthal phase factor $e^{i\ell\phi}$.
In the paraxial vortex-beam limit, $\ell$ is the usual OAM-mode label.
For vectorial or nonparaxial fields, it should be understood more generally as an azimuthal-mode index.
For any field-channel-resolved quantity $\mathcal O^{(\sigma,\ell)}$, we denote the corresponding dichroic difference by $\Delta_\zeta\mathcal O$, with $\zeta\in\{\mathrm{CD},\mathrm{HD},\mathrm{HCD}\}$.

%/_/_/_/_/_/_/_/_/_/_
\subsection{Linear structured-light absorption}
\label{sec.linearabs}
%/_/_/_/_/_/_/_/_/_/_

We start from the minimal-coupling interaction
\begin{equation}
\hat H_{\mathrm{int}}(t)
=
-
\int d\bm r\,
\hat{\bm j}(\bm r)\cdot \hat{\bm A}(\bm r,t),
\label{eq:Hint_minimal}
\end{equation}
which keeps the spatial structure of the optical mode explicit and gives a current-based description of structured-field absorption \cite{List2015JCP,Foglia2022JCP,Bonafe2025PRB}.
For a prescribed monochromatic mode, the absorption spectrum takes the nonlocal bilinear form
\begin{align}
S(\omega)
&=
\int d\bm r\, d\bm r'\,
\sum_{ab}
A_a^*(\bm r,\omega)\,
\mathcal J_{ab}(\bm r,\bm r';\omega)\,
A_b(\bm r',\omega),
\label{eq:S_linear_nonlocal}
\end{align}
where the effective absorption kernel is
\begin{equation}
\begin{split}
\mathcal J_{ab}(\bm r,\bm r';\omega)
&\equiv
-\frac{\omega}{2}\,
\mathrm{Im}
\int_{0}^{\infty} dt\,
e^{i\omega t}
\\
&\qquad\times
\left[
-\frac{i}{\hbar}
\big\langle
[\hat j_a(\bm r,t),\hat j_b(\bm r',0)]
\big\rangle_0
\right].
\end{split}
\label{eq:J_linear_def}
\end{equation}
Thus structured-light absorption is the projection of a two-point current response onto the optical bilinear
$A_a^*(\bm r,\omega)A_b(\bm r',\omega)$.

In Eq.~(\ref{eq:S_linear_nonlocal}), we work in the Coulomb (radiation) gauge for the transverse incident optical field and take $\bm A(\bm r,\omega)$ to represent the incident optical mode in this fixed gauge.
Accordingly, the absorption formula in Eq.~(\ref{eq:S_linear_nonlocal}) describes the current--vector-potential contribution within this representation.
The analysis below classifies the symmetry, tensor, and angular-mode structure of this current--field bilinear.
A fully gauge-invariant electromagnetic-response formulation would require the corresponding scalar-potential and charge-density couplings, or an explicitly field-based formulation constrained by charge conservation.
Such a reformulation is beyond the scope of the present symmetry analysis.

For a single azimuthal-mode channel, we use the paraxial representative
\begin{equation}
\bm A^{(\sigma,\ell)}(\bm r,\omega)
=
f^{(\ell)}(\rho,z;\omega)\,
e^{i\ell\phi}\,
\bm e_\sigma,
\label{eq:A_sigmaell_pure}
\end{equation}
with
$\bm e_\sigma=(\bm e_x+i\sigma\bm e_y)/\sqrt{2}$.
Equation~(\ref{eq:A_sigmaell_pure}) is introduced to retain the leading transverse structure relevant to the present symmetry and angular-channel analysis; it is not intended to represent a complete Maxwell-consistent beam solution.
A full vectorial beam in the same gauge may contain additional components, including components along the nominal propagation direction and nonparaxial corrections.
These can be incorporated in the same bilinear framework by replacing Eq.~(\ref{eq:A_sigmaell_pure}) with the corresponding Maxwell-consistent vector potential.
For the paraxial representative in Eq.~(\ref{eq:A_sigmaell_pure}), the absorption selects a diagonal angular channel of the nonlocal response.

%/_/_/_/_/_/_/_/_/_/_
\subsection{Extension to nonlinear structured-light response}
\label{sec.nonlinearabs}
%/_/_/_/_/_/_/_/_/_/_

The same nonlocal mode-space viewpoint extends to nonlinear response.
The induced current may be expanded as a functional series in the structured field \cite{Mukamel1995Book,Chernyak2015JCP},
$\langle \hat j_a(1)\rangle=\sum_{n=1}^{\infty}\langle \hat j_a(1)\rangle^{(n)}$, with
\begin{align}
\langle \hat j_a(1)\rangle^{(n)}
&=
\frac{1}{n!}
\sum_{a_1\cdots a_n}
\int d2\cdots d(n+1)\,
\notag\\
&\quad\times
R^{(n)}_{a;a_1\cdots a_n}(1;2,\ldots,n+1)
\prod_{m=1}^{n} A_{a_m}(m+1).
\label{eq:nonlinear_hierarchy_compact}
\end{align}
Here $m\equiv(\bm r_m,t_m)$, and
$R^{(n)}_{a;a_1\cdots a_n}(1;2,\ldots,n+1)$ is the $n$th-order retarded nonlocal current-response kernel.
It describes the causal response of the current at point $1$ to vector-potential insertions at points $2,\ldots,n+1$.
Schematically,
\begin{equation}
R^{(n)}
\sim
\left(-\frac{i}{\hbar}\right)^{n}
\Theta_n\,
\mathcal C^{(n+1)},
\label{eq:Rn_schematic}
\end{equation}
where $\Theta_n$ enforces causal time ordering and $\mathcal C^{(n+1)}$ denotes the nested-commutator expectation value of $n+1$ current operators.
The factor $1/n!$ corresponds to using a symmetrized $n$th-order kernel, and the linear response is recovered at $n=1$.

A scalar optical signal is obtained by contracting the induced current with a driving or detected field.
For a monochromatic detected channel $(\sigma,\ell)$, the corresponding $(n+1)$th-order contribution may be written as
\begin{align}
S^{(n+1)}_{(\sigma,\ell)}(\omega)
&=
-\frac{\omega}{2}\,
\mathrm{Im}
\int d\bm r\,
\sum_a
A^{(\sigma,\ell)*}_{a}(\bm r,\omega)\,
\langle \hat j_{a}(\bm r,\omega)\rangle^{(n)}_{(\sigma,\ell)} .
\label{eq:Sn1_eta}
\end{align}
This expression is the nonlinear analogue of Eq.~(\ref{eq:S_linear_nonlocal}).
More general nonlinear measurements may involve several driving and detection channels, but their symmetry and angular-mode content can be analyzed by the same nonlocal mode-space decomposition.

%/_/_/_/_/_/_/_/_/_/_/_/_/_/_/_/_/_/_/_/_/_/_/_
\section{Structure of dichroic response}
\label{sec.dichroism-str}
%/_/_/_/_/_/_/_/_/_/_/_/_/_/_/_/_/_/_/_/_/_/_/_

%/_/_/_/_/_/_/_/_/_/_
\subsection{Reversal-odd projection of the nonlocal response}
\label{sec.reversal}
%/_/_/_/_/_/_/_/_/_/_

We now use the absorption formula of Sec.~\ref{sec.minimal} to define the response selected by a field-reversal channel.
For a beam labeled by $(\sigma,\ell)$, the dichroic signal is
\begin{equation}
\Delta_{\zeta}S^{(\sigma,\ell)}(\omega)
=
\int d\bm r\, d\bm r'
\sum_{ab}
\Delta_{\zeta}\mathcal A_{ab}^{(\sigma,\ell)}(\bm r,\bm r')
\,
\mathcal J_{ab}(\bm r,\bm r';\omega),
\label{eq:HD_kernel}
\end{equation}
where
\begin{equation}
\Delta_{\zeta}\mathcal A_{ab}^{(\sigma,\ell)}(\bm r,\bm r')
\equiv
\Delta_{\zeta}
\left[
A_a^{*}(\bm r)A_b(\bm r')
\right]
\label{eq:dichroic_bilinear_def}
\end{equation}
is the optical-bilinear difference for the reversal channel
$\zeta\in\{\mathrm{CD},\mathrm{HD},\mathrm{HCD}\}$.
We use the unnormalized difference instead of an asymmetry ratio, so that Eq.~(\ref{eq:HD_kernel}) directly projects the reversal-odd component of the nonlocal response without mixing it with the reversal-even absorption background.

Equation~(\ref{eq:HD_kernel}) is the basic projection formula used below.
The optical field determines the kernel
$\Delta_{\zeta}\mathcal A_{ab}^{(\sigma,\ell)}$, and the measured contrast is the part of
$\mathcal J_{ab}$ that survives contraction with this reversal-odd optical bilinear.
Different reversal channels therefore select different combinations of spatial, polarization, and tensor structures of the nonlocal current response.

The symmetry interpretation depends on the reversal.
For HCD, a simple inversion-odd interpretation is valid only when the two compared complete vector fields are related by spatial inversion.
HD, by contrast, is an operational reversal of the azimuthal-mode index at fixed polarization, comparing fields with opposite transverse phase winding.
It should not by itself be identified with spatial inversion, mirror reversal, or a pseudoscalar measure of material chirality.

%/_/_/_/_/_/_/_/_/_/_
\subsection{Irreducible-tensor decomposition}
\label{sec.irreducible}
%/_/_/_/_/_/_/_/_/_/_

We decompose the nonlocal response kernel and the dichroic optical bilinear in Eq.~(\ref{eq:HD_kernel}) into irreducible tensor sectors.
This separates the structured-light signal into scalar $(J=0)$, axial-vector $(J=1)$, and rank-2 tensor $(J=2)$ response channels.

For two vectors $\bm X$ and $\bm Y$, the dyadic product $X_aY_b$ decomposes as
\begin{align}
X_aY_b
&=
\frac{1}{3}\delta_{ab}\,(\bm X\!\cdot\!\bm Y)
+
\frac{1}{2}\sum_c \epsilon_{abc}\,(\bm X\!\times\!\bm Y)_c
\nonumber\\
&\quad+
\left[
\frac{1}{2}(X_aY_b+X_bY_a)
-
\frac{1}{3}\delta_{ab}(\bm X\!\cdot\!\bm Y)
\right].
\label{eq:dyadic_decomp}
\end{align}
The three terms are the scalar, antisymmetric axial-vector, and symmetric traceless rank-2 parts, respectively.

We apply this decomposition to the response kernel $\mathcal J_{ab}$ and to the dichroic optical bilinear
\[
\Delta_{\zeta}\mathcal A_{ab}(\bm r,\bm r')
\equiv
\Delta_{\zeta}\!\left[A_a^*(\bm r)A_b(\bm r')\right].
\]
The response kernel is written as
\begin{align}
\mathcal J^{(0)}_{ab}
&=
\frac{1}{3}\delta_{ab}\mathcal J_{cc},
\label{eq:J0}
\\
\mathcal J^{(1)}_{ab}
&=
\frac{1}{2}\left(\mathcal J_{ab}-\mathcal J_{ba}\right),
\label{eq:J1}
\\
\mathcal J^{(2)}_{ab}
&=
\frac{1}{2}\left(\mathcal J_{ab}+\mathcal J_{ba}\right)
-
\frac{1}{3}\delta_{ab}\mathcal J_{cc}.
\label{eq:J2}
\end{align}
Here and below, the arguments $(\bm r,\bm r';\omega)$ are suppressed when no ambiguity arises.
The optical bilinear is decomposed in the same way:
\begin{align}
\Delta_{\zeta}\mathcal A^{(0)}_{ab}
&=
\frac{1}{3}\delta_{ab}\,
\Delta_{\zeta}\!\left[\bm A^*(\bm r)\!\cdot\!\bm A(\bm r')\right],
\label{eq:A0}
\\
\Delta_{\zeta}\mathcal A^{(1)}_{ab}
&=
\frac{1}{2}
\left(
\Delta_{\zeta}\mathcal A_{ab}
-
\Delta_{\zeta}\mathcal A_{ba}
\right),
\label{eq:A1}
\\
\Delta_{\zeta}\mathcal A^{(2)}_{ab}
&=
\frac{1}{2}
\left(
\Delta_{\zeta}\mathcal A_{ab}
+
\Delta_{\zeta}\mathcal A_{ba}
\right)
-
\frac{1}{3}\delta_{ab}\,
\Delta_{\zeta}\!\left[\bm A^*(\bm r)\!\cdot\!\bm A(\bm r')\right].
\label{eq:A2}
\end{align}

The contribution from each tensor sector is
\begin{equation}
\mathcal I^{(J)}_{\zeta}(\bm r,\bm r';\omega)
=
\mathcal J^{(J)}_{ab}(\bm r,\bm r';\omega)
\Delta_{\zeta}\mathcal A^{(J)}_{ab}(\bm r,\bm r'),
\label{eq:IJ_compact}
\end{equation}
and the integrand in Eq.~(\ref{eq:HD_kernel}) becomes
\begin{align}
\Delta_{\zeta}\mathcal A_{ab}(\bm r,\bm r')\,
\mathcal J_{ab}(\bm r,\bm r';\omega)
&=
\sum_{J=0}^{2}
\mathcal I^{(J)}_{\zeta}(\bm r,\bm r';\omega).
\label{eq:tensor_channels}
\end{align}
Equivalently, the full dichroic signal is
\begin{equation}
\Delta_{\zeta}S(\omega)
=
\sum_{J=0}^{2}
\int d\bm r\, d\bm r'\,
\mathcal I^{(J)}_{\zeta}(\bm r,\bm r';\omega).
\label{eq:HD_tensor_sum}
\end{equation}

This tensor-channel decomposition is used below.
The scalar sector is selected by the dichroic part of
$\bm A^*(\bm r)\!\cdot\!\bm A(\bm r')$,
the axial-vector sector by the antisymmetric optical bilinear, and the rank-2 sector by the symmetric traceless optical bilinear.
Thus a structured-light dichroic signal is resolved into microscopic response channels, not treated as a single pseudoscalar.

The axial-vector response $\mathcal J^{(1)}$ is most directly connected to conventional molecular optical activity because it represents antisymmetric circulating-current response.
The scalar and rank-2 sectors, $\mathcal J^{(0)}$ and $\mathcal J^{(2)}$, describe scalar spatial-dispersion and anisotropic nonlocal response.
They are not pseudoscalar chirality measures by themselves, but can contribute to reversal-odd structured-light signals when combined with inversion-odd spatial dependence and nontrivial optical mode structure.
The decomposition therefore identifies which microscopic tensor sector of the nonlocal response is selected by a given optical reversal channel.

%/_/_/_/_/_/_/_/_/_/_
\subsection{Single-mode and mixed-mode structured-light dichroism}
\label{sec.singlemode}
%/_/_/_/_/_/_/_/_/_/_

We now apply the tensor decomposition to fields with definite or mixed azimuthal-mode content.
Here ``diagonal'' means response within a single OAM-mode channel, whereas ``off-diagonal'' means interference between different OAM-mode channels.
The optical mode structure then selects both the angular-channel sector and the tensor sector of the nonlocal response.

We first consider a single azimuthal-mode channel in Eq.~(\ref{eq:A_sigmaell_pure}).
For an ordinary vortex beam, we assume
$f^{(\ell)}(\rho,z;\omega)=f^{(-\ell)}(\rho,z;\omega)$ and define
\begin{equation}
\mathcal{F}^{(\ell,\ell')}(\rho,z;\rho',z';\omega)
=
f^{(\ell)*}(\rho,z;\omega)\,
f^{(\ell')}(\rho',z';\omega).
\label{eq:helical_bilinear}
\end{equation}
The pure-mode optical bilinear is proportional to
$\mathcal F^{(\ell,\ell)}e^{i\ell(\phi'-\phi)}$.
Its azimuthal dependence therefore involves only the relative angle $\phi'-\phi$, so a single azimuthal-mode beam probes a diagonal angular channel of the nonlocal response.

The tensor content of each reversal channel then follows directly from the decomposition above.
For a circularly polarized single mode, CD is purely axial-vector.
HD and HCD contain scalar, axial-vector, and rank-2 sectors, but with different angular dependence:
in HD all three sectors are proportional to $\sin[\ell(\phi'-\phi)]$, whereas in HCD the axial-vector sector is proportional to $\cos[\ell(\phi'-\phi)]$ while the scalar and rank-2 sectors remain sine-type.
For linearly polarized single-mode fields, the axial-vector HD sector is absent, leaving only scalar and rank-2 sectors.
The explicit tensor forms are given in Appendix~\ref{app:singlemode_tensor_sectors}.

We next introduce a controlled mixed-mode field,
\begin{equation}
\bm A(\bm r,\omega)
=
\bm A^{(\sigma,\ell)}(\bm r,\omega)
+
\alpha\,
\bm A^{(\sigma',0)}(\bm r,\omega),
\label{eq:A_mix}
\end{equation}
with real mixing amplitude $\alpha$. Here and below, polarization reversal of the mixed field means \((\sigma,\sigma')\to(-\sigma,-\sigma')\).
Because absorption is bilinear in the field, the new terms are the interference bilinears
\begin{align}
\mathcal A_{ab}^{(\sigma,\ell;\sigma',0)}(\bm r,\bm r';\omega)
&=
\alpha e_{\sigma,a}^*e_{\sigma',b}\,
\mathcal F^{(\ell,0)}(\rho,z;\rho',z';\omega)\,
e^{-i\ell\phi},
\label{eq:offdiag_helical_bilinear_1}
\\
\mathcal A_{ab}^{(\sigma',0;\sigma,\ell)}(\bm r,\bm r';\omega)
&=
\alpha e_{\sigma',a}^*e_{\sigma,b}\,
\mathcal F^{(0,\ell)}(\rho,z;\rho',z';\omega)\,
e^{+i\ell\phi'} .
\label{eq:offdiag_helical_bilinear_2}
\end{align}
These phase factors depend separately on $\phi$ and $\phi'$, not only on $\phi'-\phi$.
They therefore probe off-diagonal angular-channel components of the nonlocal kernel connecting the $\ell$ and $0$ channels.
This is the main distinction between pure-mode and mixed-mode structured-light dichroism.

The relative polarization of the two components controls the tensor sector of this off-diagonal response.
When $\sigma=\sigma'$, the mixed contribution retains the same hierarchy as the single-mode case: mixed CD is purely axial-vector, while mixed HD and HCD generally contain all three tensor sectors.
When $\sigma=-\sigma'$, the scalar and axial-vector sectors vanish and only a rank-2 off-diagonal channel remains.
Thus the OAM admixture selects the off-diagonal angular channel, while the relative polarization selects its tensor character.
The opposite-polarization case is also a natural setting for polarization textures that can support optical skyrmions~\cite{ChengEtAl2026AOP}.
The explicit mixed-mode tensor forms are listed in Appendix~\ref{app:mixedmode_tensor_sectors}.

The same distinction extends to nonlinear response.
In the linear signal of Eq.~(\ref{eq:HD_kernel}), the mixed contribution probes a single off-diagonal $\ell$--$0$ channel of the two-point kernel.
In the $(n+1)$th-order signal of Eq.~(\ref{eq:Sn1_eta}), the detected field is contracted with an $n$th-order nonlocal current induced by $n$ field insertions [Eq.~(\ref{eq:nonlinear_hierarchy_compact})].
For azimuthal components $A^{(m)}\propto e^{im\phi}$, continuous axial symmetry requires conservation of the total azimuthal mode index in the measured scalar signal.
Including the material angular channel $\mu^{(n)}$ carried by the response kernel, the selection rule is
$m_{\mathrm{det}}-\sum_{j=1}^{n}m_j+\mu^{(n)}=0$,
up to the sign convention for the detected channel.
Under a discrete $C_N$ symmetry, the same condition holds modulo $N$.
Thus nonlinear response can make mixed-mode coherence observable through angular-momentum matching beyond the linear $\ell$--$0$ channel.

%/_/_/_/_/_/_/_/_/_/_
\subsection{Mode-space interpretation and gradient limit}
\label{sec.gradientlim}
%/_/_/_/_/_/_/_/_/_/_

The two-point formulation clarifies how local-gradient descriptions emerge from the nonlocal current response.
Introducing midpoint and relative coordinates,
$\bm R=(\bm r+\bm r')/2$ and $\bm\delta=\bm r'-\bm r$,
the optical bilinear can be expanded as
\begin{align}
A_a^{*}(\bm r)A_b(\bm r')
&\approx
A_a^{*}(\bm R)A_b(\bm R)
+\frac{\delta_i}{2}
\Bigl[
A_a^{*}(\bm R)\partial_iA_b(\bm R)
\nonumber\\
&\qquad\qquad
-
(\partial_iA_a^{*}(\bm R))A_b(\bm R)
\Bigr]
+\cdots .
\label{eq:bilinear_gradient}
\end{align}
The angular and tensor structures of the nonlocal optical bilinear are then converted into local field amplitudes and gradients.
Thus the present theory provides a parent two-point expression for local multipolar, spatial-dispersion, and optical-chirality descriptions when the field and the effective current response vary slowly on the relevant microscopic scale.

For a single azimuthal-mode channel, the gradient expansion gives local counterparts of the three tensor sectors.
The scalar sector corresponds to symmetric spatial-dispersion response.
It is sensitive to local phase gradients, including the azimuthal phase variation of a vortex beam, and should be viewed as a local marker of inversion-odd nonlocal response, not as a pointwise chirality measure.

The axial-vector sector gives the familiar chiroptical structure.
For example, it generates antisymmetric polarization-gradient combinations such as
\begin{equation}
E_x^{*}\partial_z E_y - E_y^{*}\partial_z E_x,
\label{eq:optical_chirality_schematic}
\end{equation}
with all fields evaluated at $\bm R$.
Such terms have the same character as optical-chirality structures associated with Lipkin's zilch and related formulations \cite{Lipkin1964JMP,TangCohen2010PRL}.
In the present language, they are the local-gradient form of the $J=1$ circulating-current channel.

The rank-2 sector gives the tensorial extension of this local picture.
At the gradient level it is governed by the symmetric traceless field anisotropy,
\begin{align}
\mathrm{Im}\!\left[
E_a^{*}\partial_i E_b
+
E_b^{*}\partial_i E_a
\right]
-\frac{2}{3}\delta_{ab}\,
\mathrm{Im}\!\left[
E_c^{*}\partial_i E_c
\right].
\label{eq:Q_gradient_tensor_E}
\end{align}
This channel describes anisotropic and quadrupole-like distortion of the structured field and is naturally connected to tensor-based descriptions of electromagnetic chirality \cite{SmaginDyakov2026arXiv}.
Together with the scalar sector, it shows that the local-gradient limit contains more information than the axial-vector optical-chirality channel alone.

The local-gradient viewpoint is useful when the two-point response can be represented by low-order spatial moments.
This includes short-range current response, as in metallic nanostructures and metasurface elements \cite{Ni2021ACSNano,Jain2024ACSNano}, and smooth electronic states whose wave functions or transition-current densities vary slowly on the optical-mode scale.
In such cases, the gradient expansion provides a practical route to local scalar, axial-vector, and rank-2 response channels, while local optical chirality and related near-field chirality densities are interpreted as gradient-level projections of the underlying nonlocal current response \cite{TangCohen2010PRL}.

For mixed modes, the local-gradient form retains the mode-space interference in Eqs.~(\ref{eq:offdiag_helical_bilinear_1}) and (\ref{eq:offdiag_helical_bilinear_2}), and therefore still carries information about off-diagonal angular-channel coupling.
This connection is interpretive: the fundamental objects remain the nonlocal optical bilinears and their contraction with the nonlocal current-response kernel.
Thus the local-gradient limit provides a controlled bridge to local multipolar and spatial-dispersion descriptions, while the underlying formulation remains the nonlocal two-point response.

%/_/_/_/_/_/_/_/_/_/_
\subsection{From symmetry selection to diagnostic analysis}
\label{sec.low_symmetry_analysis}
%/_/_/_/_/_/_/_/_/_/_

The tensor and mode-space decomposition developed above turns the symmetry analysis into a diagnostic scheme for structured-light dichroism.
Local optical chirality and related densities provide useful field-side descriptors in the local-gradient regime.
The present framework keeps this local viewpoint, but asks a more diagnostic question: which material response is selected by a given structured-light contrast.
The measured signal is resolved by mode composition, polarization, and position, so that structured-light dichroism can probe nonlocal material response beyond local optical chirality.

This viewpoint is useful when ideal symmetry arguments no longer give a simple selection rule.
Material symmetries restrict the allowed reversal channels and angular-channel couplings, as summarized in Appendix~\ref{app:symmetry_selection}.
In finite nanostructures, metasurfaces, oriented molecular assemblies, and thin films, boundaries, substrates, disorder, beam displacement, and orientational anisotropy can relax these restrictions.
A finite HD- or HCD-type signal should then be analyzed as a channel-resolved response, not assigned to chirality alone.

The first diagnostic step is to separate diagonal and off-diagonal angular-channel response.
A pure azimuthal-mode beam, such as Eq.~(\ref{eq:A_sigmaell_pure}), mainly probes the diagonal response of the nonlocal kernel.
Adding a controlled zero-OAM component, as in Eq.~(\ref{eq:A_mix}), introduces $\ell$--$0$ interference and tests whether the observed contrast comes from a single OAM channel or from coupling between different angular channels.
For a polarization configuration $\mathcal P$, the admixture dependence may be fitted as
\begin{equation}
\Delta_\zeta S_{\mathrm{mix}}(\alpha;\mathcal P)
=
\Delta_\zeta S_{\mathrm{diag}}(\mathcal P)
+
\alpha\,D_\zeta(\mathcal P)
+
O(\alpha^2).
\label{eq:analysis_alpha_expansion}
\end{equation}
The intercept gives the diagonal single-channel contribution, and the slope $D_\zeta(\mathcal P)$ gives the component linear in the $\ell$--$0$ interference.
Changing the sign of $\alpha$ or comparing several small values of $\alpha$ checks that this component is a linear interference term, not an intensity correction.
Thus $D_\zeta(\mathcal P)$ is the experimentally extracted off-diagonal signal to be analyzed next.

The second diagnostic step is to characterize the tensor content accessible from the polarization dependence within the present measurement geometry.
For linear polarization at angle $\beta$, the slope can be expanded as
\begin{equation}
D_\zeta(\beta)
=
D_\zeta^{(0)}
+
D_{\zeta,c}^{(2)}\cos 2\beta
+
D_{\zeta,s}^{(2)}\sin 2\beta .
\label{eq:analysis_tensor_tomography_linear}
\end{equation}
The angle-independent coefficient $D_\zeta^{(0)}$ denotes the response that is invariant under in-plane polarization rotation, whereas $D_{\zeta,c}^{(2)}$ and $D_{\zeta,s}^{(2)}$ characterize the angle-dependent in-plane anisotropy.
Because the scan in Eq.~(\ref{eq:analysis_tensor_tomography_linear}) is restricted to the $xy$ plane, however, $D_\zeta^{(0)}$ need not be purely scalar.
An axially symmetric rank-2 contribution with equal in-plane diagonal components and a distinct out-of-plane component is also angle independent and can contribute to $D_\zeta^{(0)}$.
Distinguishing such a contribution from the scalar response requires additional information from a measurement involving the out-of-plane direction.
The circular-polarization-odd contribution is obtained separately by comparing the two circular-polarization labels.
Writing $D_{\zeta,\sigma}\equiv D_\zeta(\mathcal P_\sigma)$ for the slope measured with circular polarization $\sigma=\pm1$, we define
\begin{equation}
D_\zeta^{(1)}
=
\frac{1}{2}
\left(
D_{\zeta,+}
-
D_{\zeta,-}
\right),
\label{eq:analysis_tensor_tomography_axial}
\end{equation}
up to the normalization convention for the circular basis.
Here $\mathcal P_\sigma$ denotes the polarization configuration based on $\bm e_\sigma=(\bm e_x+i\sigma\bm e_y)/\sqrt{2}$.
The coefficient $D_\zeta^{(1)}$ measures the circular-polarization-odd part of the material nonlocal current response and reflects the material response selected by the corresponding axial optical bilinear rather than the optical chirality of the probe field itself.
Taken together, $D_\zeta^{(0)}$, $D_\zeta^{(1)}$, and $(D_{\zeta,c}^{(2)},D_{\zeta,s}^{(2)})$ indicate whether the mixed-channel contrast is governed mainly by an angle-independent in-plane response, a circular-current-type response, or angle-dependent in-plane anisotropy.
The opposite-circular-polarization mixed mode discussed in Sec.~\ref{sec.singlemode} provides a direct anisotropic-response filter because the scalar and axial-vector sectors vanish in that configuration.

The extracted channels can then be related to their spatial and symmetry-breaking origins.
A beam-position map $\Delta_\zeta S(\bm R_0)$ shows whether a given channel is localized at edges, corners, defects, substrate-modified regions, or optical hot spots.
Changing $\ell$ tests which angular-channel mismatch couples to the material response and, when an approximate $C_N$ symmetry is present, checks the corresponding modulo-$N$ selection rule.
Sample rotation gives the complementary angular analysis,
$\Delta_\zeta S(\theta_s)=\sum_m C_m^{(\zeta)}e^{im\theta_s}$,
where $C_m^{(\zeta)}$ labels the material angular harmonics selected by the optical channel.
In the local-gradient limit, these position-dependent signals can be interpreted through local optical chirality, spatial dispersion, and tensorial anisotropy.

This diagnostic scheme clarifies what structured-light dichroism can reveal in low-symmetry systems.
The $\alpha$-scan separates diagonal and off-diagonal angular response, polarization scans characterize the tensor content accessible within the measurement geometry, and spatial or angular scans locate the symmetry-breaking source.
Thus a finite HD- or HCD-type contrast becomes a mode-, tensor-, and position-resolved probe of nonlocal optical response.

%/_/_/_/_/_/_/_/_/_/_
\subsection{Relation to representative experiments}
\label{sec.exper}
%/_/_/_/_/_/_/_/_/_/_

Representative HD-type experiments can be interpreted through this diagnostic viewpoint.
For molecular and solution-phase measurements \cite{Jain2023JCP}, the OAM-reversal channel compares absorption for the $\ell$ and $-\ell$ modes under otherwise fixed polarization conditions.
Orientational averaging suppresses many sample-specific anisotropic contributions, so the measured contrast primarily reflects the reversal-odd part of the nonlocal response extracted from the much larger reversal-even absorption background \cite{KatoYokoshi2024PRB}.
Controlled variations of beam alignment, mode composition, or weak zero-OAM admixture would provide additional information on the balance between diagonal OAM-channel response and off-diagonal mode interference.

Nanophotonic and metasurface platforms provide a natural setting for the full diagnostic scheme.
In single chiral nanostructures \cite{Ni2021ACSNano}, a sizable HD-type signal may indicate that an inversion-odd response is already active in a predominantly diagonal single-mode channel.
Beam-position maps can then reveal where the response is concentrated, such as edges, tips, or other symmetry-breaking parts of the structure.
In plasmonic metasurfaces and structured beams with finite mode mixing \cite{Jain2024ACSNano}, reduced symmetry and nonideal mode composition make the mixed-mode viewpoint particularly useful.
The $\alpha$- and polarization-dependences discussed in Sec.~\ref{sec.low_symmetry_analysis} then provide diagnostic handles on the mode-space and tensor character of the observed contrast.

Nonlinear molecular experiments \cite{Begin2023NatPhoton} lie outside the linear absorption regime.
The present linear theory does not aim at a quantitative description of those measurements, but the same mode-space viewpoint remains useful for organizing the relevant angular-momentum channels.
A full comparison requires an extension to nonlinear current response together with a microscopic model of the material and optical fields.

%/_/_/_/_/_/_/_/_/_/_/_/_/_/_/_/_/_/_/_/_/_/_/_
\section{Conclusion}
%/_/_/_/_/_/_/_/_/_/_/_/_/_/_/_/_/_/_/_/_/_/_/_

We have developed a microscopic framework for structured-light dichroism based on the nonlocal current response.
A dichroic signal is formulated as a reversal-odd, channel-resolved projection of the nonlocal response kernel onto the optical bilinear of the incident field.
This viewpoint separates symmetry selection, tensor character, and angular-mode structure within one description.

The central distinction is between diagonal and off-diagonal angular-channel response.
A single azimuthal-mode field probes diagonal OAM-channel components of the nonlocal kernel.
A mixed field introduces interference between different OAM-mode channels and thereby probes off-diagonal mode-space coherence.
The polarization composition of the interfering components then selects the tensor character of this response.
In particular, opposite circular polarizations isolate a rank-2 off-diagonal channel, giving a direct route to tensorial structured-light dichroism.

This mode-space viewpoint also clarifies the role of symmetry.
In high-symmetry systems, inversion, mirror, and rotational symmetries impose selection rules on the allowed reversal channels and angular-mode couplings.
In low-symmetry nanophotonic structures, finite systems, and extended materials, these restrictions can be relaxed and several channels may coexist.
The present decomposition then provides a diagnostic procedure: mode composition separates diagonal and off-diagonal angular response, polarization characterizes the tensor content accessible within the measurement geometry, and spatial or angular scans locate the source of symmetry breaking.

The same nonlocal formulation connects structured-light dichroism to local descriptions.
Expanding the two-point optical bilinear gives local-gradient structures associated with spatial dispersion, optical chirality, and tensorial anisotropy.
Thus local chirality densities and related gradient quantities appear as limiting forms of the underlying nonlocal current response, not as independent starting assumptions.
Because the optical mode profile is retained explicitly, the framework can be extended to nonlinear response, vectorial and nonparaxial beams, and multipolar regimes.

These results position structured-light dichroism as a symmetry-, mode-, and tensor-resolved probe of nonlocal optical response.
The theory provides a bridge between microscopic current-response descriptions, local chiroptical quantities, and experimentally tunable structured-light channels in molecular, nanophotonic, and extended material systems.

%/_/_/_/_/_/_/_/_/_/_/_/_/_/_/_/_/_/_/_/_/_/_/_
\section*{Acknowledgement}
%/_/_/_/_/_/_/_/_/_/_/_/_/_/_/_/_/_/_/_/_/_/_/_
This work was supported by the JSPS KAKENHI Grants Nos.~JP21H05019, JP22K04863, JP22H05131 and JP22H05132, and by the JSPS International Joint Research Program JRP-LEAD with UKRI under Grant No. JPJSJRP20241710. This work was also supported by JST ERATO Grant No. JPMJER2503.

%/_/_/_/_/_/_/_/_/_/_/_/_/_/_/_/_/_/_/_/_/_/_/_/_/_/_/_/_/_/_/_/_/_/_/_/_/_/_/_
\section*{Data Availability}
%/_/_/_/_/_/_/_/_/_/_/_/_/_/_/_/_/_/_/_/_/_/_/_/_/_/_/_/_/_/_/_/_/_/_/_/_/_/_/_

No data were created or analyzed in this study.

\appendix

%/_/_/_/_/_/_/_/_/_/_/_/_/_/_/_/_/_/_/_/_/_/_/_
\section{Explicit single-mode tensor sectors}
\label{app:singlemode_tensor_sectors}
%/_/_/_/_/_/_/_/_/_/_/_/_/_/_/_/_/_/_/_/_/_/_/_

We list the explicit tensor forms of the single-mode optical sectors discussed in Sec.~\ref{sec.singlemode}.
These expressions make the tensorial content of the dichroic optical bilinears fully explicit and show the main-text statements in Cartesian form.
In the single-mode case, all sectors remain diagonal in angular-channel space and depend only on the relative angle $\phi'-\phi$.
All expressions below carry the common factor $\mathcal F^{(\ell,\ell)}(\rho,z;\rho',z';\omega)$.

For CD, defined by $\sigma\to-\sigma$ at fixed $\ell$,
\begin{align}
\Delta_{\mathrm{CD}}\mathcal A_{ab}^{(0)(\sigma,\ell)}
&=0,
\label{eq:CD_A0}
\\
\Delta_{\mathrm{CD}}\mathcal A_{ab}^{(1)(\sigma,\ell)}
&=
i\sigma\,\epsilon_{abz}\,
e^{i\ell(\phi'-\phi)},
\label{eq:CD_A1}
\\
\Delta_{\mathrm{CD}}\mathcal A_{ab}^{(2)(\sigma,\ell)}
&=0.
\label{eq:CD_A2}
\end{align}
Thus, in the single-mode case, CD is purely axial-vector.

For HD, defined here as $\ell\to-\ell$ at fixed $\sigma$,
\begin{align}
\Delta_{\mathrm{HD}}\mathcal A_{ab}^{(0)(\sigma,\ell)}
&=
\frac{2i}{3}\delta_{ab}\,
\sin[\ell(\phi'-\phi)],
\label{eq:HD_A0}
\\
\Delta_{\mathrm{HD}}\mathcal A_{ab}^{(1)(\sigma,\ell)}
&=
-\sigma\,\epsilon_{abz}\,
\sin[\ell(\phi'-\phi)],
\label{eq:HD_A1}
\\
\Delta_{\mathrm{HD}}\mathcal A_{ab}^{(2)(\sigma,\ell)}
&=
\frac{i}{3}\,
\sin[\ell(\phi'-\phi)]
\nonumber\\
&\quad\times
\left(
\delta_{ax}\delta_{bx}
+
\delta_{ay}\delta_{by}
-
2\delta_{az}\delta_{bz}
\right).
\label{eq:HD_A2}
\end{align}
Hence the HD-type OAM-reversal channel involves all three tensor sectors, all with the same sine-type angular dependence.
For comparison, in the corresponding single-mode case with linear polarization, the axial-vector sector vanishes, and only the scalar and rank-2 sectors remain.

For HCD, defined by $(\sigma,\ell)\to(-\sigma,-\ell)$,
\begin{align}
\Delta_{\mathrm{HCD}}\mathcal A_{ab}^{(0)(\sigma,\ell)}
&=
\frac{2i}{3}\delta_{ab}\,
\sin[\ell(\phi'-\phi)],
\label{eq:HCD_A0}
\\
\Delta_{\mathrm{HCD}}\mathcal A_{ab}^{(1)(\sigma,\ell)}
&=
i\sigma\,\epsilon_{abz}\,
\cos[\ell(\phi'-\phi)],
\label{eq:HCD_A1}
\\
\Delta_{\mathrm{HCD}}\mathcal A_{ab}^{(2)(\sigma,\ell)}
&=
\frac{i}{3}\,
\sin[\ell(\phi'-\phi)]
\nonumber\\
&\quad\times
\left(
\delta_{ax}\delta_{bx}
+
\delta_{ay}\delta_{by}
-
2\delta_{az}\delta_{bz}
\right).
\label{eq:HCD_A2}
\end{align}
HCD also contains all three sectors, but with a characteristic separation: the scalar and rank-2 sectors are sine-type, whereas the axial-vector sector is cosine-type.
For a linearly polarized single azimuthal-mode channel, this distinction disappears because the axial-vector sector is absent, and the HD and HCD bilinear structures coincide.

%/_/_/_/_/_/_/_/_/_/_/_/_/_/_/_/_/_/_/_/_/_/_/_
\section{Explicit mixed-mode tensor sectors}
\label{app:mixedmode_tensor_sectors}
%/_/_/_/_/_/_/_/_/_/_/_/_/_/_/_/_/_/_/_/_/_/_/_

We next list the explicit tensor forms for the mixed-mode case discussed in Sec.~\ref{sec.singlemode}.
The key feature is interference between the $\ell$ and $0$ OAM-mode channels.
Unlike the single-mode expressions above, the mixed sectors are off-diagonal in mode space and do not reduce to functions of the relative angle $\phi'-\phi$ alone.
To display this structure explicitly, we introduce the shorthand notation
\begin{align}
X_{\ell}(\bm r,\bm r';\omega)
&\equiv
\mathcal F^{(\ell,0)}(\rho,z;\rho',z';\omega)\,e^{-i\ell\phi},
\label{eq:Xell_def}
\\
Y_{\ell}(\bm r,\bm r';\omega)
&\equiv
\mathcal F^{(0,\ell)}(\rho,z;\rho',z';\omega)\,e^{+i\ell\phi'},
\label{eq:Yell_def}
\end{align}
and
\begin{equation}
\begin{split}
\Sigma_{\ell}(\bm r,\bm r';\omega)
&\equiv
X_{\ell}(\bm r,\bm r';\omega)
+
Y_{\ell}(\bm r,\bm r';\omega),
\\
\Delta_{\ell}(\bm r,\bm r';\omega)
&\equiv
X_{\ell}(\bm r,\bm r';\omega)
-
Y_{\ell}(\bm r,\bm r';\omega).
\end{split}
\label{eq:SigmaDelta_def}
\end{equation}
These combinations separate the interference into symmetric and antisymmetric parts under exchange of the two off-diagonal bilinears and make the tensor-sector structure more transparent.

For the same-polarization case $\sigma=\sigma'$, the interference bilinear becomes
\begin{equation}
\mathcal M_{ab}^{(\ell),\sigma\parallel}
=
e_{\sigma,a}^*e_{\sigma,b}\,\Sigma_{\ell},
\label{eq:M_same}
\end{equation}
so that the mixed CD, HD, and HCD sectors are
\begin{align}
\Delta_{\mathrm{CD}} \mathcal A_{ab}^{(0)\,\mathrm{(int)},\,\sigma\parallel}
&=0,
\label{eq:mix_same_CD_A0}
\\
\Delta_{\mathrm{CD}} \mathcal A_{ab}^{(1)\,\mathrm{(int)},\,\sigma\parallel}
&=
i\alpha\sigma\,\epsilon_{abz}\,\Sigma_{\ell},
\label{eq:mix_same_CD_A1}
\\
\Delta_{\mathrm{CD}} \mathcal A_{ab}^{(2)\,\mathrm{(int)},\,\sigma\parallel}
&=0.
\label{eq:mix_same_CD_A2}
\end{align}
\begin{align}
\Delta_{\mathrm{HD}} \mathcal A_{ab}^{(0)\,\mathrm{(int)},\,\sigma\parallel}
&=
\frac{\alpha}{3}\delta_{ab}\,
\bigl(\Sigma_{\ell}-\Sigma_{-\ell}\bigr),
\label{eq:mix_same_HD_A0}
\\
\Delta_{\mathrm{HD}} \mathcal A_{ab}^{(1)\,\mathrm{(int)},\,\sigma\parallel}
&=
\frac{i\alpha\sigma}{2}\epsilon_{abz}\,
\bigl(\Sigma_{\ell}-\Sigma_{-\ell}\bigr),
\label{eq:mix_same_HD_A1}
\\
\Delta_{\mathrm{HD}} \mathcal A_{ab}^{(2)\,\mathrm{(int)},\,\sigma\parallel}
&=
\frac{\alpha}{6}
\bigl(
\delta_{ax}\delta_{bx}
+\delta_{ay}\delta_{by}
-2\delta_{az}\delta_{bz}
\bigr)
\nonumber\\
&\quad\times
\bigl(\Sigma_{\ell}-\Sigma_{-\ell}\bigr).
\label{eq:mix_same_HD_A2}
\end{align}
and
\begin{align}
\Delta_{\mathrm{HCD}} \mathcal A_{ab}^{(0)\,\mathrm{(int)},\,\sigma\parallel}
&=
\frac{\alpha}{3}\delta_{ab}\,
\bigl(\Sigma_{\ell}-\Sigma_{-\ell}\bigr),
\label{eq:mix_same_HCD_A0}
\\
\Delta_{\mathrm{HCD}} \mathcal A_{ab}^{(1)\,\mathrm{(int)},\,\sigma\parallel}
&=
\frac{i\alpha\sigma}{2}\epsilon_{abz}\,
\bigl(\Sigma_{\ell}+\Sigma_{-\ell}\bigr),
\label{eq:mix_same_HCD_A1}
\\
\Delta_{\mathrm{HCD}} \mathcal A_{ab}^{(2)\,\mathrm{(int)},\,\sigma\parallel}
&=
\frac{\alpha}{6}
\bigl(
\delta_{ax}\delta_{bx}
+\delta_{ay}\delta_{by}
-2\delta_{az}\delta_{bz}
\bigr)
\nonumber\\
&\quad\times
\bigl(\Sigma_{\ell}-\Sigma_{-\ell}\bigr).
\label{eq:mix_same_HCD_A2}
\end{align}
In this case, the tensor structure remains close to the single-mode one: mixed CD stays purely axial-vector, while the mixed HD and HCD channels generally contain all three tensor sectors.

For the opposite-polarization case $\sigma=-\sigma'$, one instead has
\begin{equation}
\mathcal M_{ab}^{(\ell),\sigma\perp}
=
\frac{1}{2}
\bigl(\delta_{ax}\delta_{bx}-\delta_{ay}\delta_{by}\bigr)\Sigma_{\ell}
-
\frac{i\sigma}{2}
\bigl(\delta_{ax}\delta_{by}+\delta_{ay}\delta_{bx}\bigr)\Delta_{\ell},
\label{eq:M_opposite}
\end{equation}
so that the scalar and axial-vector sectors vanish identically and only the rank-2 sector remains.
In this case,
\begin{align}
\Delta_{\mathrm{CD}} \mathcal A_{ab}^{(0)\,\mathrm{(int)},\,\sigma\perp}
&=0,
\label{eq:mix_opp_CD_A0}
\\
\Delta_{\mathrm{CD}} \mathcal A_{ab}^{(1)\,\mathrm{(int)},\,\sigma\perp}
&=0,
\label{eq:mix_opp_CD_A1}
\\
\Delta_{\mathrm{CD}} \mathcal A_{ab}^{(2)\,\mathrm{(int)},\,\sigma\perp}
&=
-i\alpha\sigma\,
\bigl(\delta_{ax}\delta_{by}+\delta_{ay}\delta_{bx}\bigr)\,
\Delta_{\ell},
\label{eq:mix_opp_CD_A2}
\end{align}
\begin{align}
\Delta_{\mathrm{HD}} \mathcal A_{ab}^{(0)\,\mathrm{(int)},\,\sigma\perp}
&=0,
\label{eq:mix_opp_HD_A0}
\\
\Delta_{\mathrm{HD}} \mathcal A_{ab}^{(1)\,\mathrm{(int)},\,\sigma\perp}
&=0,
\label{eq:mix_opp_HD_A1}
\\
\Delta_{\mathrm{HD}} \mathcal A_{ab}^{(2)\,\mathrm{(int)},\,\sigma\perp}
&=
\frac{\alpha}{2}
\bigl(\delta_{ax}\delta_{bx}-\delta_{ay}\delta_{by}\bigr)
\bigl(\Sigma_{\ell}-\Sigma_{-\ell}\bigr)
\nonumber\\
&\quad
-
\frac{i\alpha\sigma}{2}
\bigl(\delta_{ax}\delta_{by}+\delta_{ay}\delta_{bx}\bigr)
\bigl(\Delta_{\ell}-\Delta_{-\ell}\bigr).
\label{eq:mix_opp_HD_A2}
\end{align}
and
\begin{align}
\Delta_{\mathrm{HCD}} \mathcal A_{ab}^{(0)\,\mathrm{(int)},\,\sigma\perp}
&=0,
\label{eq:mix_opp_HCD_A0}
\\
\Delta_{\mathrm{HCD}} \mathcal A_{ab}^{(1)\,\mathrm{(int)},\,\sigma\perp}
&=0,
\label{eq:mix_opp_HCD_A1}
\\
\Delta_{\mathrm{HCD}} \mathcal A_{ab}^{(2)\,\mathrm{(int)},\,\sigma\perp}
&=
\frac{\alpha}{2}
\bigl(\delta_{ax}\delta_{bx}-\delta_{ay}\delta_{by}\bigr)
\bigl(\Sigma_{\ell}-\Sigma_{-\ell}\bigr)
\nonumber\\
&\quad
-
\frac{i\alpha\sigma}{2}
\bigl(\delta_{ax}\delta_{by}+\delta_{ay}\delta_{bx}\bigr)
\bigl(\Delta_{\ell}+\Delta_{-\ell}\bigr).
\label{eq:mix_opp_HCD_A2}
\end{align}
Opposite circular polarizations isolate a purely rank-2 off-diagonal channel and make the polarization-resolved tensor structure of mixed-mode dichroism explicit.

%/_/_/_/_/_/_/_/_/_/_/_/_/_/_/_/_/_/_/_/_/_/_/_
\section{Field transformations and symmetry selection rules}
\label{app:symmetry_selection}
%/_/_/_/_/_/_/_/_/_/_/_/_/_/_/_/_/_/_/_/_/_/_/_

We summarize the symmetry constraints on the absorption functional.
This also fixes the relation between the operational reversal channels used in the main text and spatial symmetry operations of the optical field.
The statements below concern complete vector-field profiles, not label changes such as $\ell\rightarrow-\ell$ alone.

Let $g$ be a spatial symmetry operation of the material, with vector representation $D(g)$.
A polar vector field transforms as
\begin{equation}
A_a^{g}(\bm r,\omega)
=
D_{ab}(g)A_b(g^{-1}\bm r,\omega).
\label{eq:app_Ag_general}
\end{equation}
If the current-response kernel satisfies
\begin{equation}
\mathcal J_{ab}(\bm r,\bm r';\omega)
=
D_{ac}(g)D_{bd}(g)
\mathcal J_{cd}(g^{-1}\bm r,g^{-1}\bm r';\omega),
\label{eq:app_Jg_general}
\end{equation}
then Eq.~(\ref{eq:S_linear_nonlocal}) gives $S[A^g]=S[A]$.
Thus a dichroic difference vanishes when the two incident fields are related by a symmetry operation of the material.
For spatial inversion, $g=P$ and $D(P)=-I$.
Then $A_a^P(\bm r,\omega)=-A_a(-\bm r,\omega)$, while an inversion-symmetric material satisfies $\mathcal J_{ab}(\bm r,\bm r';\omega)=\mathcal J_{ab}(-\bm r,-\bm r';\omega)$ because the two polar-vector current signs cancel.
Hence two fields related by full spatial inversion have identical absorption in a centrosymmetric material.
If the combined reversal corresponds to full inversion of the complete vector field, the corresponding HCD difference vanishes in an inversion-symmetric material.

This statement is not equivalent to the label change $\ell\rightarrow-\ell$ alone.
Parity acts on the complete field, including the spatial arguments, propagation direction, polar-vector character of $\bm A$, and possible longitudinal or nonparaxial components.
Only when the parity-transformed field is represented, within a specified beam convention, by $(\sigma,\ell)\rightarrow(-\sigma,-\ell)$ does the corresponding label reversal acquire the inversion property above.
For the mirror plane $xz$, $M_y:(x,y,z)\mapsto(x,-y,z)$ and $D(M_y)=\mathrm{diag}(1,-1,1)$.
If the material is invariant under $M_y$, then $S[A^{M_y}]=S[A]$.
For the simple paraxial field in Eq.~(\ref{eq:A_sigmaell_pure}), $M_y$ sends $\phi\rightarrow-\phi$ and $D(M_y)\bm e_\sigma=\bm e_{-\sigma}$, giving the label structure $(\sigma,\ell)\rightarrow(-\sigma,-\ell)$ within this convention.
For focused, displaced, substrate-modified, or nonparaxial beams, the complete field must instead be transformed by Eq.~(\ref{eq:app_Ag_general}).

Inversion, mirror reflection, and label reversal should be distinguished.
For optical vortices, this distinction is important because the labels $\sigma$ and $\ell$ depend on the transverse phase convention and on the chosen propagation direction.
In the paraxial convention used here, mirror reflection reverses the transverse azimuthal angle and changes both $\sigma$ and $\ell$, whereas spatial inversion also reverses the propagation direction and the polar-vector character of $\bm A$.
The representation of spatial inversion in terms of $(\sigma,\ell)$ is convention dependent.
Accordingly, $\ell$-reversal at fixed $\sigma$ is treated as an operational azimuthal-mode-index reversal, not by itself as mirror reflection or spatial inversion of the complete optical field.

Rotational symmetry constrains angular-channel mixing.
For continuous axial symmetry, the kernel depends on the azimuthal angles only through their difference, \(
\mathcal J_{ab}(\rho,\phi,z;\rho',\phi',z';\omega)
=
\mathcal J_{ab}(\rho,\rho',z,z';\phi'-\phi;\omega)
\), apart from the Cartesian tensor transformation.
A single azimuthal-mode bilinear contains $e^{i\ell(\phi'-\phi)}$ and is compatible with this diagonal angular structure.
In contrast, the mixed $\ell$--$0$ bilinears in Eqs.~(\ref{eq:offdiag_helical_bilinear_1}) and (\ref{eq:offdiag_helical_bilinear_2}) contain $e^{-i\ell\phi}$ or $e^{+i\ell\phi'}$.
For example, suppressing tensor and radial variables, \(
\int d\phi\,d\phi'\,
\mathcal J_{\mathrm{env}}(\phi'-\phi)e^{-i\ell\phi}
=
\left(\int d\Phi\,e^{-i\ell\Phi}\right)
\left(\int d\varphi\,\mathcal J_{\mathrm{env}}(\varphi)\right)
\), with $\Phi=\phi$ and $\varphi=\phi'-\phi$.
This vanishes for $\ell\neq0$, and the same holds for the conjugate factor $e^{+i\ell\phi'}$.
Thus a continuously rotationally symmetric system has no linear mixed $\ell$--$0$ contribution in this scalar-envelope sector.

For a discrete $C_N$ symmetry, angular momentum is conserved modulo $N$.
Writing the scalar-envelope part of the kernel as
\begin{equation}
\mathcal J_{ab}(\phi,\phi';\omega)
=
\sum_{m,m'}
\mathcal J^{m,m'}_{ab}(\omega)
e^{im\phi}e^{-im'\phi'},
\label{eq:app_J_mmprime}
\end{equation}
gives the selection rule $m-m'\equiv0\pmod N$.
Since the mixed $\ell$--$0$ optical bilinear connects envelope channels differing by $\ell$, its scalar-envelope contribution is allowed only when $\ell\equiv0\pmod N$.
For vectorial or nonparaxial fields, this condition applies to the total angular phase of the complete optical bilinear, including polarization-dependent and longitudinal contributions.

When inversion, mirror, or rotational symmetries are lowered or absent, these restrictions are released, and off-diagonal angular-channel components can contribute to the dichroic signal, as discussed in Sec.~\ref{sec.low_symmetry_analysis}.

\end{document}